\documentclass{article}
\usepackage[T1]{fontenc} 
\usepackage[utf8]{inputenc} 

\usepackage{spconf,amsmath,cite,url}
\usepackage{graphicx}
\usepackage{color}
\usepackage{algorithm}
\usepackage[noend]{algpseudocode}
\usepackage{amsfonts}
\usepackage{amssymb}
\usepackage{listings}
\usepackage{xparse}
\usepackage{microtype}
\usepackage{booktabs}
\usepackage{tabularx}
\usepackage{units}
\title{Mind the beat: detecting audio onsets from EEG recordings of music listening}

\name{Ashvala Vinay \qquad Alexander Lerch \qquad Grace Leslie}
\address{Center for Music Technology, Georgia Institute of Technology}
 
\begin{document}

\maketitle

\begin{abstract}

We propose a deep learning approach to predicting audio event onsets in electroencephalogram (EEG) recorded from users as they listen to music. We use a publicly available dataset containing ten contemporary songs and concurrently recorded EEG. We generate a sequence of onset labels for the songs in our dataset and trained neural networks (a fully connected network (FCN) and a recurrent neural network (RNN)) to parse one second windows of input EEG to predict one second windows of onsets in the audio. We compare our RNN network to both the standard spectral-flux based novelty function and the FCN. We find that our RNN was able to produce results that reflected its ability to generalize better than the other methods. 

Since there are no pre-existing works on this topic, the numbers presented in this paper may serve as useful benchmarks for future approaches to this research problem.

\end{abstract}

\begin{keywords}
EEG, MIR, Deep Learning, Audio Onset detection
\end{keywords}


\section{Introduction}\label{sec:introduction}

\begin{figure}
 \centerline{
 \includegraphics[width=\columnwidth]{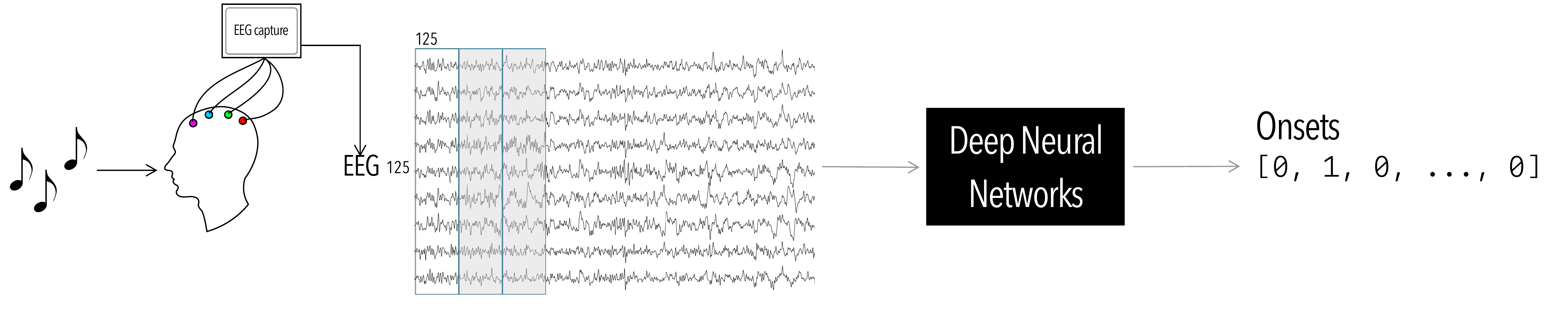}}
 \caption{Given EEG recording during music listening, we propose applying neural networks to the prediction of audio onsets in the music.}
 \label{fig:example}
\end{figure}

Our work seeks to measure the degree to which we may solve the problem: how precisely can we extract audio onsets from the music a person is listening to, given access only the user's EEG signal? A visual representation of the problem is shown in Fig.~\ref{fig:example}. 

A primary goal of music imagery information retrieval (MIIR) research is to "recognize the music in our thoughts \cite{Stober2015}." Researchers in this nascent field have sought to reconstruct audio from electroencephalogram (EEG) recordings of music listening, known to be an intractable problem due to the indirect, epiphenomenal relationship between sensory input and the resulting EEG traces measured from a listener, in addition to the presence of confounding artifacts in the EEG data itself. 

In contrast, onset detection is a well-defined problem in the field of music information retrieval (MIR) in which an onset detection function (ODF) is developed to identify the onset of an acoustic event in a given recording of music. Contemporary solutions to the problem rely on deep neural networks (DNNs) \cite{eyben2010universal, schluter2013musical}. 

We propose that current solutions to the problem of audio onset estimation may lay the foundation for a solution to the more difficult task of estimating audio onsets from EEG data. A solution to this problem would enable new technologies that predict music features from a user's EEG signal allowing the user to control, for example, music recommendation systems or generative systems for music composition and performance. While our title refers to the musical concept of beats, our paper focuses on the more fundamental underlying process of beat or onset detection. 

%

In the upcoming section, we review past research in computational EEG analysis, MIR and MIIR that lend methods we apply to the present research problem. Section~\ref{sec:methods} outlines our present problem and defines the recurrent neural network (RNN) architectures we evaluate in our experiments and our overall evaluation strategy. Section~\ref{sec:results} is dedicated to results and discussion. Section~\ref{sec:conclusion} discusses future research directions.


\section{Related Work}\label{sec:relwork}


Hans Berger \cite{PMID:16334737} invented EEG as a diagnostic tool in understanding human brain activity. His first EEG recordings were instrumental in understanding the difference between a normal and an abnormal brain. Since then, there has been an interest in understanding of the connection between an auditory stimulus and the resulting EEG trace it elicits. This is a lively area of research in neuroscience.  The earliest research was completed in clinical settings in which doctors attempted to better understand the human auditory system by observing rhythmic fluctuations in voltage measured from the human scalp, notably in patients experiencing musicogenic epilepsy \cite{doi:10.1111/j.1600-0447.1949.tb07349.x}. An early non-diagnostic use of EEG and music comes from Walker \cite{walker1980alpha}, who used EEG to measure performance on a music recognition task. 

More recently, researchers in the field of music cognition have designed controlled experiments aimed at defining the features of EEG that shift predictably in the presence of rhythmic musical tones \cite{fujioka2015beta, heilbron2018great}. For example, Fujioka et al.\ modeled audio features in physiological responses by using alternating loud and soft tones at a fixed \unit[390]{ms} inter-onset interval as stimulus \cite{Fujioka}. They used wavelet transforms to decompose data into a time-frequency representation to better analyze synchronicity between onsets in audio and onsets in the MEG data. In the MEG signal, they found spikes in activity in the Beta (\unit[10--25]{Hz}) and Gamma (\unit[28--48]{Hz}) bands at intervals consistent with the inter-onset interval of the audio. Fujioka et al. follow up their work in 2015 \cite{fujioka2015beta} where they investigate synchronous presence of beats of specific music styles in the beta bands of MEG they recorded in response to music stimuli. Heilbron and Chait \cite{heilbron2018great} explore the possibility of predictive coding in the auditory cortex of the brain given that there is evidence of time locking provided above. Their work opens the doors for us to investigate research questions that seek to understand feature representation of music in the brain.

In contrast to these well-controlled auditory experiments, more naturalistic experimental approaches have focused on gathering physiological data from music listeners as they enjoy the music they listen to on a regular basis. For instance, one recent study collected EEG recordings of 48 participants listening to natural music stimuli to understand how participants enjoyed music based on the modifications to a song \cite{KANESHIRO2020116559}. In a similar example, the researchers analyzed fMRI of music listening to understand music preferences, predict the genre a person might be thinking of, and identify the song \cite{Casey2017}.

In parallel to this new vein of naturalistic listening research, there has been a significant shift towards machine learning and deep learning based approaches that may be applied to feature extraction and information retrieval from these naturalistic music stimuli \cite{StoberISMIR:2016, DBLP:journals/corr/StoberSOG15}. These techniques have also been applied to research questions combining music and EEG.

Ofner and Stober influenced the present research direction by using EEG data of music and speech listening to generate  the mel-spectrogram of the EEG and audio \cite{Ofner2018}. Specifically, they used one-second windows of raw EEG as input to a neural network which generates one second of mel-spectrogram of music as output. Their work highlights the potential of using neural networks for MIIR specific tasks. It is among the first works to make use of a dataset called NMED-T \cite{Losorelli2017} for machine learning tasks. However, due to the lack of definition in the evaluation procedure and metrics, their results remain inconclusive.

In addition to NMED-T, other multi-modal EEG-audio datasets have been released in recent years which enable data-driven approaches to human auditory research. Of these, the OpenMIIR dataset \cite{Stober2015} studies the difference between imagined and perceived music as represented by the EEG of the listener. Researchers at CCRMA, Stanford, have released datasets in addition to the previously mentioned NMED-T: NMED-RP studies the perception of rhythm \cite{appaji2018neural}, whereas NMED-H \cite{dmochowskilate} studies the perception of full-length popular Bollywood works. 

The field of music information retrieval has contributed important methods that we draw upon in this research. The extraction of onsets from audio signals has long been a core problem in MIR, feeding into systems such as tempo detection or music transcription. Onsets mark the beginning of an acoustic event \cite{lerch2012introduction}, and there have been multiple contributions to detecting onsets in a music signal. Two state-of-the-art methods use deep learning methods to predict onsets and use spectrogram representations of the music signal as inputs \cite{eyben2010universal,schluter2013musical}. While Schlüter and Böck use a convolutional neural network to perform the same task, Eyben et al.\ use bi-directional RNNs to model onsets. Conventional RNNs use layers of hidden states to process inputs across time. Both methods are implemented in the MADMOM library \cite{madmom}. We use MADMOM in our work to generate pseudo-ground truth annotations upon which our networks are trained.

\section{Experimental Setup}\label{sec:methods}

We extract onsets in music from the EEG signals recorded during music listening sessions. Since this problem is an information retrieval problem, we adopt approaches used in the MIR and MIIR domains. Our work takes cues from Ofner and Stober \cite{Ofner2018}; their work influenced our decision on handling EEG inputs and our choice of dataset. In this section, we discuss the experiments undertaken to evaluate the feasibility of an EEG audio onset detection solution: how we handle the flow of data and evaluate the output of our networks.

\subsection{Dataset and data representation}

We use the NMED-T dataset \cite{Losorelli2017}. The dataset is a collection of EEG recordings from subjects listening to contemporary alternative music. The dataset contains ten songs and EEG from twenty participants. All songs are between four and five minutes long. The EEG data is available in both a raw 128-channel and a pre-processed 125-channel format. We use the latter variant, which is provided at a sample rate of \unit[125]{Hz}. Each file provided in the pre-processed version provides us with a $ 20 \times 125 \times N $ matrix (20 subjects, 125 channels and $N$ time steps).

\subsubsection{Audio onset generation}

In order to evaluate any EEG audio onset detection method, ground truth audio onset data is required as a basis for comparison. As the NMED-T dataset does not provide such onset data for the music stimuli, this ground truth onset information needs to be annotated either by hand or with an automated detection method. The amount of data precludes hand-annotation; since we want the most robust set of annotations for onsets, we choose to work with MADMOM's RNN ODF \cite{eyben2010universal} to generate our ground truth onset information. MADMOM contains the state-of-the-art robust music onset detection methods introduced in Sect.~\ref{sec:relwork}, which outperform other onset detection methods \cite{eyben2010universal, schluter2013musical}. 

The present method is expected to have a \unit[7]{\%} error rate for a given song \cite{eyben2010universal}, producing a reliable set of annotations that we can use as our pseudo-ground truth. MADMOM's ODFs take an audio file as input and output a series of timestamps. We use these timestamps to generate a binary sequence with a sample rate of \unit[125]{Hz} where $1$ indicates the presence of an onset at the timestep and $0$ indicates the absence of an onset. 

\subsubsection{Pre-processing} 

EEG data are typically preprocessed using a standard set of functions designed to limit the influence of noise and artifacts. In this case, we filter the EEG data with a bandpass filter (\unit[0.1--40]{Hz}) and zero-pad the dataset uniformly to a length of $37500$ samples (\unit[5]{min}), since the longest song in the dataset is around five minutes long. The resulting EEG signals are segmented into one second blocks as suggested by Ofner and Stober \cite{Ofner2018}. 

The NMED-T dataset is provided in a song-wise format: each file represents both a song and the 20 subjects' EEG recording while listening to that song. We rearrange the data to sequentially reflect a subject-wise format, where each file would instead represent a subject's perception of all songs. As the dataset is considerably smaller than other datasets currently used in both the MIR and image processing domains, k-fold cross-validation (with $ k = 20 $ for the number of subjects) is used to train and validate our models.
 
\begin{figure}
 \centerline{
 \includegraphics[width=0.95\columnwidth]{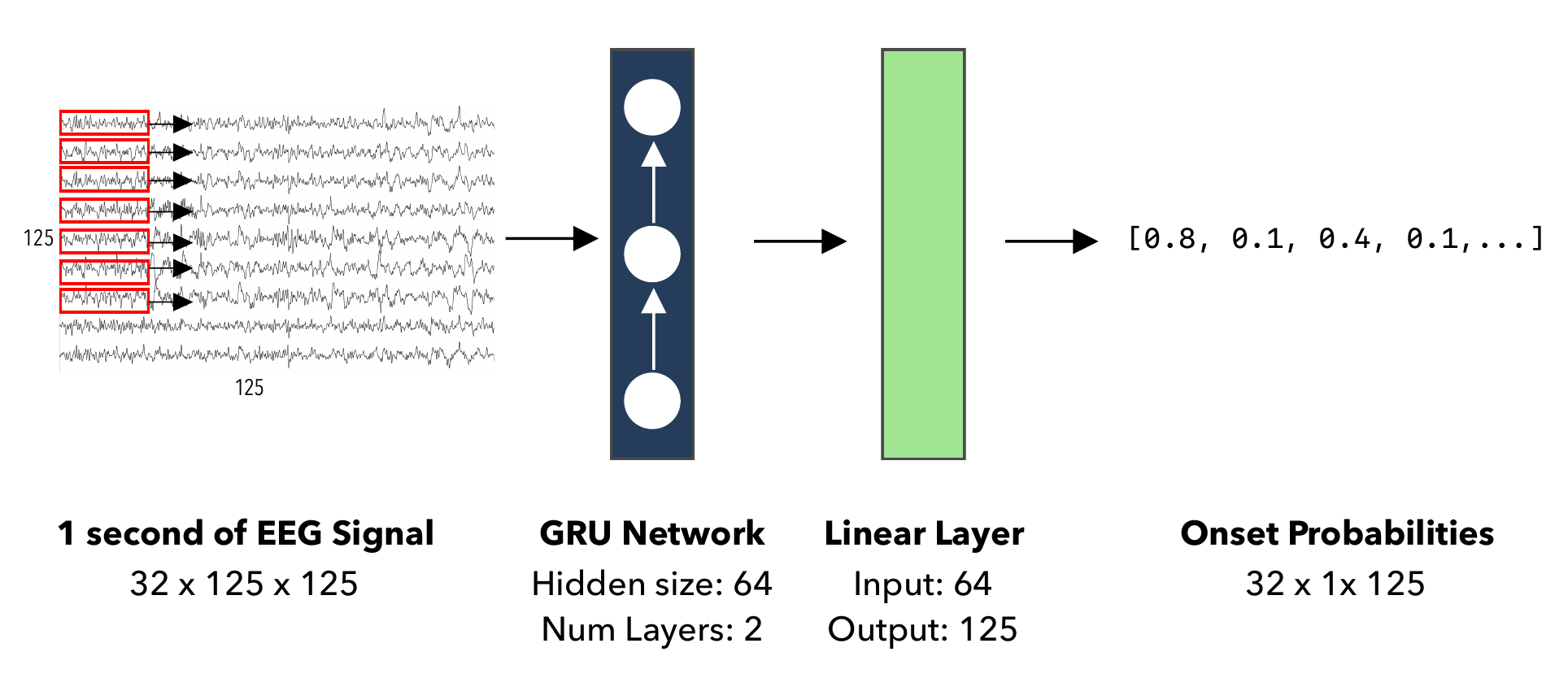}}
 \caption{Our RNN architecture predicts onsets in music using an EEG signal of music listening as input.}
 \label{fig}
\end{figure}

\subsection{Architectures}

\subsubsection{Fully connected network}

We choose a fully connected network (FCN) as a baseline comparison network for its simple architectural design and convert the 125 channels $ \times $ 125 time-step signal into a vector of the  dimension  $1 \times 15625$. The output is a $ 1 \times 125 $ sequence which represents the equivalent binary onset output sequence in the audio example. We use a two-layer FCN with a hidden layer of size 256. 

\subsubsection{Recurrent Neural Network}

RNNs have been shown to be suitable for modeling both sequences and time-series \cite{connor1994recurrent}. We use gated recurrent units (GRUs) as our RNN architecture of choice, given its similarity to a long-short term memory (LSTM) network. LSTMs are effective at learning long-term dependencies and retaining context across time using gates \cite{hochreiter1997long}. The choice of GRUs is more suitable to our task since the small amount of data requires a comparably small number of training parameters to avoid over-fitting. 
A two layer GRU with a hidden state size of 64 predicts onsets using a $125 \times 125$ EEG input (or one second of EEG). The output of the GRU is fed into a fully connected layer, which generates a $ 1 \times 125 $ sequence containing onset probabilities. This architecture is shown in Figure~\ref{fig}.

\subsection{Spectral Flux} 

We wanted to understand how well a standard method of detecting onsets in audio would operate when using EEG data. In order to do this, we used a spectral flux-based novelty function built into the \texttt{pyACA} package \cite{lerch2012introduction}. The pyACA ODF takes in a music signal and returns timestamps in seconds for onsets. Since we have a $ 125 $-channel EEG input, we extract estimated onset timestamps per channel and average them before feeding our evaluation pipeline.

\subsection{Training} 

Our dataset is presented to the model in the form of $ \{e_i, a_i\} $, where $ e_i $ is the $ 125 \times 125 $ matrix with 125 channels of EEG data for 1 second of data (or 125 timesteps) and $ a_i $ is a $ 1 \times 125 $ binary vector indicating the presence of onsets at a given time step. We use binary cross-entropy as our loss function to train our network.

During training, the networks output a $ 1 \times 125 $ sequence of logits. The loss function internally applies a sigmoid to the logits to convert them into a series of log-probabilities indicating the probability of an onset at a given timestep before computing cross entropy. During evaluation, a sigmoid is applied to the output of the network directly. We train our models for a total of 50 epochs each with the Adam optimizer and a fixed learning rate of $1\times10^{-3}$.

\subsection{Evaluation}

The model outputs are assessed with \texttt{mir\_eval} toolkit's onset evaluation method \cite{raffel2014mir_eval}. The onset evaluation method compares the timestamps of the ground truth with the prediction and reports the F-measure, precision and recall scores. Traditionally, onset detection models employ a peak picking on the generated output to identify positions of the most-likely onsets. We use the approach proposed by Böck et al.~\cite{bock2012evaluating}, which uses a series of tunable parameters $\operatorname{w_1} \dots \operatorname{w_5}$ and a fixed threshold $ \delta $ for difference in computed average values. Their algorithm for peak picking is usable for both real-time peak picking uses and asynchronous tasks such as ours. 

The series of selected onsets is mapped to timestamps for evaluation. In order to investigate the timing accuracy, these time steps are evaluated with several tolerance window sizes. Onset detection systems are usually evaluated with tolerance windows between 50 and \unit[100]{ms}  \cite{eyben2010universal, schluter2013musical}. We present results for tolerance windows of length \unit[0.05]{s}, \unit[0.1]{s}, \unit[0.15]{s}, \unit[0.25]{s}, \unit[0.5]{s}, \unit[0.75]{s}, \unit[1]{s}, and \unit[2]{s}. Increasing the window size provides more room for correct predictions in our output. For instance, a tolerance of \unit[50]{ms} means that an onset will be classified as being 'correct' if it is within 50 milliseconds of an annotated onset. Additionally, as we are using cross validation, all results are evaluated out of fold. 

\begin{figure}
\centering
 \includegraphics[width=0.95\columnwidth]{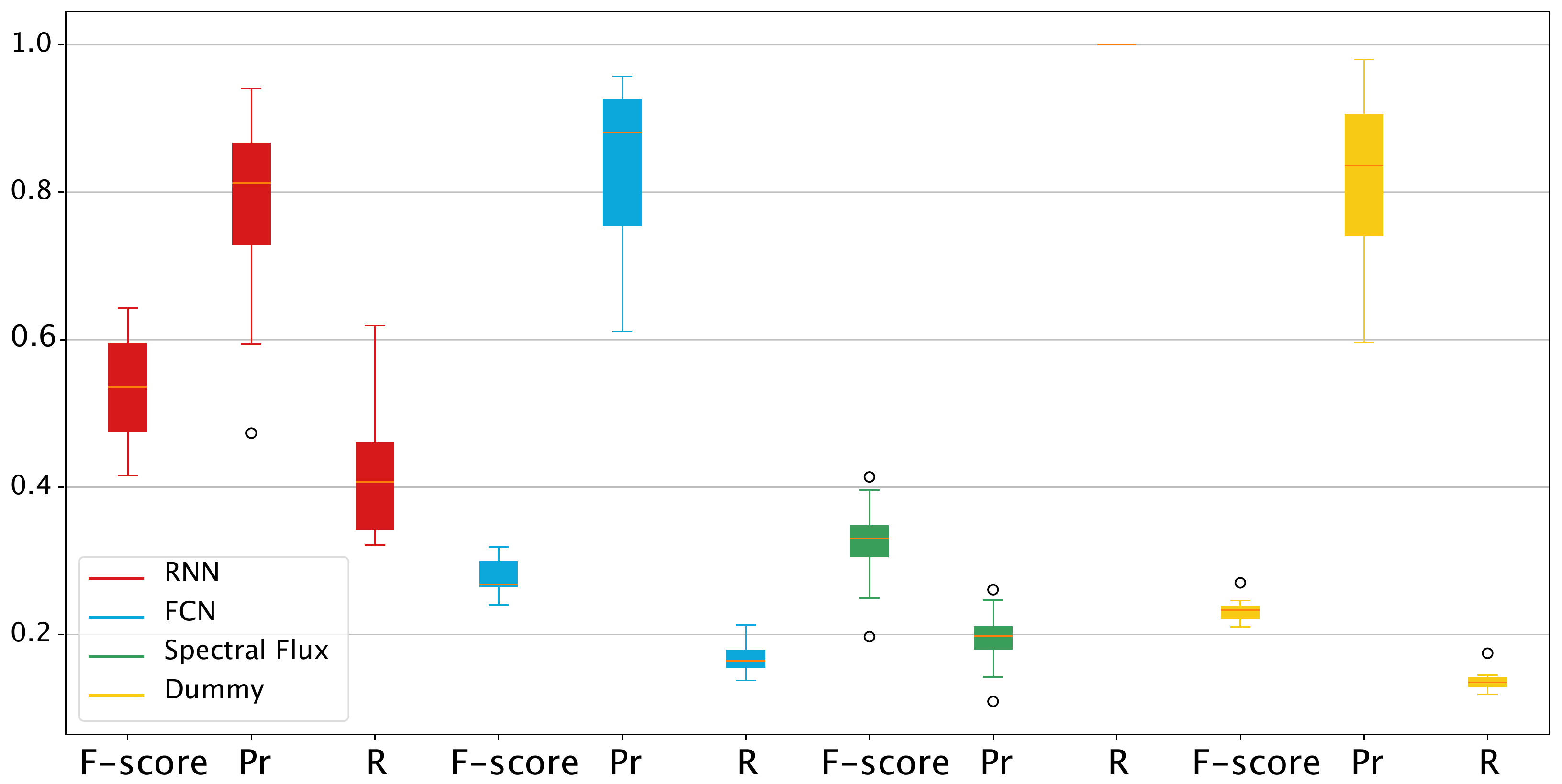}
 \caption{Each box plot represents a metric evaluated through the onset evaluation toolkit. The comparisons are done with the RNN, FCN, Spectral Flux methods and a dummy method. The results shown here are evaluated at a tolerance window of \unit[100]{ms}. The RNN network produces a higher F-score across all songs and showcases room for improvement. } 
 \label{fig:metrics}
\end{figure}

\section{Results and Discussion}\label{sec:results}

Figure~\ref{fig:metrics} shows the difference in performance between the two networks we trained as revealed by F-measures, precision and recall metrics, all evaluated at a \unit[100]{ms} tolerance window. Overall, the RNN outperforms the FCN. While the FCN performs well as measured by precision, it displays poor recall performance; this means that the FCN produced a high number of false negatives. On the other hand, the RNN produced similar precision, but was able to generate fewer false negatives, resulting in a considerably better F-measure. We implemented a dummy method which produces an onset every second. This method performed as well as the FCN, further highlighting the poor performance of the network architecture.

\begin{figure}
\centering
 \includegraphics[width=0.95\columnwidth]{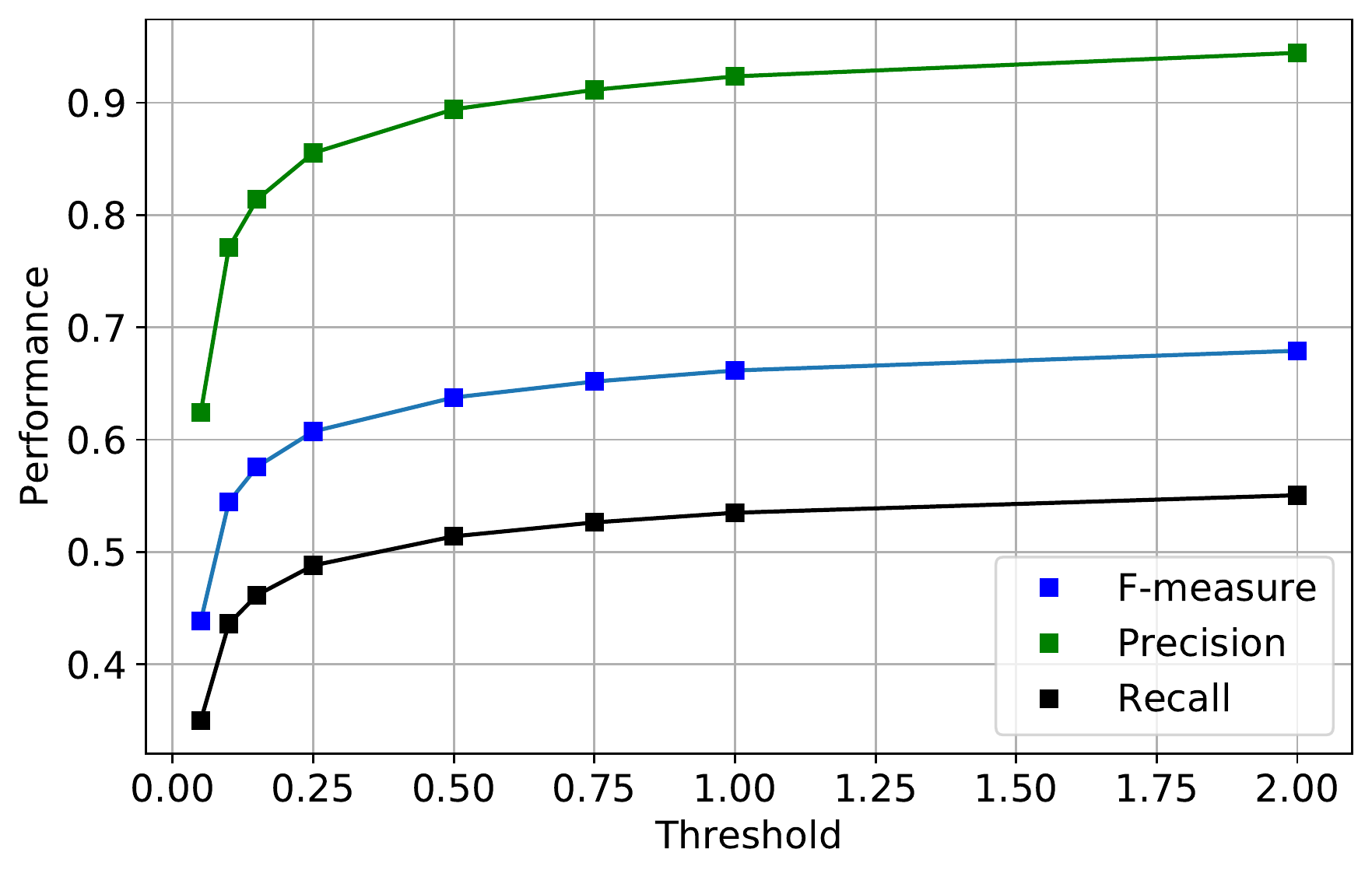}
 \caption{Tolerance windows and their impact on performance across each performance metric. Most gains are realized by the \unit[750]{ms} mark.}
  \label{fig:thresh}
\end{figure}

Fig.~\ref{fig:thresh} shows the results as a function of the tolerance window length. Increasing the tolerance window size leads, as expected, to better results across the board for all metrics. We observed that the increased windows increased precision greatly, and had less impact on the recall. This indicates that our network is producing a higher number of false negatives than desired. The gains start to converge at \unit[250]{ms} and do not change much for windows sizes larger than \unit[750]{ms}. This indicates a reduced time accuracy compared to audio-based onset detection methods which can be explained by the more noisy and complex input data.

Simply applying traditional music oriented methods to EEG data did not perform well on this task --- peak-picking a spectral flux-based novelty function yielded an average F-measure of $0.32$ (see Fig.~\ref{fig:metrics}), emphasizing the power of the proposed method. However, compared to models extracting onsets from the audio data, the EEG onset detection is considerably less accurate. While we can expect an average F-measure of nearly \unit[90]{\%} for audio data \cite{schluter2013musical}, our method can only achieve \unit[54]{\%}.
This large difference can obviously be attributed to the much more noisy and indirect EEG input data; however, it is a very encouraging result that our approach can detect onsets from EEG data with a reasonable accuracy. 

Across all subjects, we found that the average F-scores were similar to the reported scores in Fig.~\ref{fig:metrics} ($ \mu $ = 0.416, $\sigma$ = 0.08), indicating a reasonable amount of inter-subject variability. Additionally, we performed a correlation analysis between the  RNN results, and the metadata provided by NMED-T and found no significant correlations between the reported F-measures and age, musical training, listening habits.

\section{Conclusion}\label{sec:conclusion}
This research asks the question: how well can we predict onsets in music using physiological data recorded in response to that music? Simply applying music-oriented ODFs to EEG data did not work well; they require music signals sampled at audio rates and cannot generalize to physiological data at lower sample rates. Therefore, we developed an RNN architecture and experimental setup to extract onsets in music using EEG. Our encouraging results demonstrate the feasibility of constructing and testing a network to extract onsets in music, providing a basis for comparison for future studies. 

Our results also encourage future research in reconstructing other features and creating a path for being able to reconstruct stimuli found in the EEG signal. For onset modeling, we believe that the future will rely on the creation of a dataset specifically intended to evaluate the performance of a network in extracting onsets from EEG of music listening. 


\bibliographystyle{IEEEbib}
\bibliography{ICASSP}

\begin{thebibliography}{10}

\bibitem{Stober2015}
Sebastian Stober, Avital Sternin, Adrian~M Owen, and Jessica~A Grahn,
\newblock ``{Towards Music Imagery Information Retrieval: Introducing the
  OpenMIIR Dataset of EEG Recordings from Music Perception and Imagination},''
\newblock {\em 16th International Society for Music Information Retrieval
  Conference (ISMIR'15)}, 2015.

\bibitem{eyben2010universal}
Florian Eyben, Sebastian B{\"o}ck, Bj{\"o}rn Schuller, and Alex Graves,
\newblock ``Universal onset detection with bidirectional long-short term memory
  neural networks,''
\newblock in {\em Proc. 11th Intern. Soc. for Music Information Retrieval
  Conference, ISMIR, Utrecht, The Netherlands}, 2010, pp. 589--594.

\bibitem{schluter2013musical}
Jan Schl{\"u}ter and Sebastian B{\"o}ck,
\newblock ``Musical onset detection with convolutional neural networks,''
\newblock in {\em 6th international workshop on machine learning and music
  (MML), Prague, Czech Republic}, 2013.

\bibitem{PMID:16334737}
Mario Tudor, Lorainne Tudor, and Katarina~Ivana Tudor,
\newblock ``[hans berger (1873-1941)--the history of electroencephalography],''
\newblock {\em Acta medica Croatica : casopis Hravatske akademije medicinskih
  znanosti}, vol. 59, no. 4, pp. 307—313, 2005.

\bibitem{doi:10.1111/j.1600-0447.1949.tb07349.x}
H.~P.~Stubbe Teglbjærg,
\newblock ``On musciogenic epilepsy,''
\newblock {\em Acta Psychiatrica Scandinavica}, vol. 24, no. 3‐4, pp.
  679--690, 1949.

\bibitem{walker1980alpha}
James~L Walker,
\newblock ``Alpha eeg correlates of performance on a music recognition task,''
\newblock {\em Physiological Psychology}, vol. 8, no. 3, pp. 417--420, 1a980.

\bibitem{fujioka2015beta}
Takako Fujioka, Bernhard Ross, and Laurel~J Trainor,
\newblock ``Beta-band oscillations represent auditory beat and its metrical
  hierarchy in perception and imagery,''
\newblock {\em Journal of Neuroscience}, vol. 35, no. 45, pp. 15187--15198,
  2015.

\bibitem{heilbron2018great}
Micha Heilbron and Maria Chait,
\newblock ``Great expectations: is there evidence for predictive coding in
  auditory cortex?,''
\newblock {\em Neuroscience}, vol. 389, pp. 54--73, 2018.

\bibitem{Fujioka}
Takako Fujioka, Laurel~J. Trainor, Edward~W. Large, and Bernhard Ross,
\newblock ``Beta and gamma rhythms in human auditory cortex during musical beat
  processing,''
\newblock {\em Annals of the New York Academy of Sciences}, vol. 1169, no. 1,
  pp. 89--92, 2009.

\bibitem{KANESHIRO2020116559}
Blair Kaneshiro, Duc~T. Nguyen, Anthony~M. Norcia, Jacek~P. Dmochowski, and
  Jonathan Berger,
\newblock ``Natural music evokes correlated eeg responses reflecting temporal
  structure and beat,''
\newblock {\em NeuroImage}, vol. 214, pp. 116559, 2020.

\bibitem{Casey2017}
Michael~A. Casey,
\newblock ``{Music of the 7Ts: Predicting and decoding multivoxel fMRI
  responses with acoustic, schematic, and categorical music features},''
\newblock {\em Frontiers in Psychology}, vol. 8, no. JUL, pp. 1--11, 2017.

\bibitem{humphrey2012moving}
Eric~J Humphrey, Juan~Pablo Bello, and Yann LeCun,
\newblock ``Moving beyond feature design: Deep architectures and automatic
  feature learning in music informatics.,''
\newblock in {\em ISMIR}. Citeseer, 2012, pp. 403--408.

\bibitem{StoberISMIR:2016}
Sebastian Stober, Thomas Pratzlich, and Meinard Muller,
\newblock ``{Brain Beats: Tempo Extraction from EEG Data},''
\newblock in {\em Proceedings of the 17th International Society for Music
  Information Retrieval Conference}, New York, NY, aug 2016.

\bibitem{Ofner2018}
Andr{\'{e}} Ofner and Sebastian Stober,
\newblock ``{Shared Generative Representation of Auditory Concepts and EEG to
  Reconstruct Perceived and Imagined Music},''
\newblock {\em 19th International Society for Music Information Retrieval
  Conference – ISMIR 2018}, pp. 392--399, 2018.

\bibitem{Losorelli2017NMEDTAT}
Steven Losorelli, Duc~T. Nguyen, Jacek Dmochowski, and Blair Kaneshiro,
\newblock ``Nmed-t: A tempo-focused dataset of cortical and behavioral
  responses to naturalistic music,''
\newblock in {\em ISMIR}, 2017.

\bibitem{appaji2018neural}
Jay Appaji and Blair Kaneshiro,
\newblock ``Neural tracking of simple and complex rhythms: Pilot study and
  dataset,''
\newblock {\em Late-Breaking Demos Session for ISMIR}, 2018.

\bibitem{dmochowskilate}
Anthony M~Norcia Dmochowski, Jonathan Berger, and Naturalistic Music~EEG
  Dataset—Hindi,
\newblock ``Late-breaking/demo,''
\newblock .

\bibitem{lerch2012introduction}
Alexander Lerch,
\newblock {\em An introduction to audio content analysis: Applications in
  signal processing and music informatics},
\newblock Wiley-IEEE Press, 2012.

\bibitem{madmom}
Sebastian B{\"o}ck, Filip Korzeniowski, Jan Schl{\"u}ter, Florian Krebs, and
  Gerhard Widmer,
\newblock ``{madmom: a new Python Audio and Music Signal Processing Library},''
\newblock in {\em Proceedings of the 24th ACM International Conference on
  Multimedia}, Amsterdam, The Netherlands, 10 2016, pp. 1174--1178.

\bibitem{connor1994recurrent}
Jerome~T Connor, R~Douglas Martin, and Les~E Atlas,
\newblock ``Recurrent neural networks and robust time series prediction,''
\newblock {\em IEEE transactions on neural networks}, vol. 5, no. 2, pp.
  240--254, 1994.

\bibitem{hochreiter1997long}
Sepp Hochreiter and J{\"u}rgen Schmidhuber,
\newblock ``Long short-term memory,''
\newblock {\em Neural computation}, vol. 9, no. 8, pp. 1735--1780, 1997.

\bibitem{raffel2014mir_eval}
Colin Raffel, Brian McFee, Eric~J Humphrey, Justin Salamon, Oriol Nieto, Dawen
  Liang, Daniel~PW Ellis, and C~Colin Raffel,
\newblock ``mir\_eval: A transparent implementation of common mir metrics,''
\newblock in {\em In Proceedings of the 15th International Society for Music
  Information Retrieval Conference, ISMIR}. Citeseer, 2014.

\bibitem{bock2012evaluating}
Sebastian B{\"o}ck, Florian Krebs, and Markus Schedl,
\newblock ``Evaluating the online capabilities of onset detection methods.,''
\newblock in {\em ISMIR}, 2012, pp. 49--54.

\end{thebibliography}

\end{document}